\documentclass{appolb}
\usepackage{graphicx,amssymb}

\begin{document}
\title{Unified quark-hadron EoS and critical endpoint in the QCD phase diagram
\thanks{Presented at the Workshop on "Criticality in QCD and the Hadron Resonance Gas",
Wroclaw, 31. July 2020}%
}
\author{David Blaschke
\address{Institute of Theoretical Physics, University Wroc{\l}aw, Poland\\
Bogoliubov Laboratory for Theoretical Physics, JINR Dubna, Russia\\
National Research Nuclear University (MEPhI) Moscow, Russia}
}
\maketitle
\begin{abstract}
We present a recent development towards a unified description of quark-hadron matter in the QCD phase diagram that is based on a cluster decomposition of the generalized Beth-Uhlenbeck approach to quark matter, selfconsistently coupled to Polyakov-loop and mesonic background fields. 
The Mott dissociation of hadrons under extreme conditions of temperature and density is triggered by chiral symmetry restoration and confining aspects are modeled by the coupling to the background mean fields. 
First results for the QCD phase diagram with the capability to describe critical endpoints as well as crossover-all-over are presented and an excellent agreement with lattice QCD on the temperature axis is obtained.
\end{abstract}
\PACS{12.38.Mh, 25.75.Nq,97.60.Jd}
  
\section{Introduction}
The QCD phase diagram is a key to understanding the critical behavior of QCD at finite temperature $T$ and chemical potential
$\mu$. 
The well established limiting cases, however, the hadron resonance gas (HRG) and the perturbative quark-gluon plasma (pQGP), fail to be applicable just in the relevant hadron-to-quark matter transition region. 
Here a nonperturbative approach is required that has to adequately describe at least these key phenomena
\begin{enumerate}
\item bound state formation and dissociation, 
\item chiral symmetry breaking and restoration,
\item confinement and deconfinement of quarks and gluons.
\end{enumerate}
The only nonperturbative approach that addresses these effects ab-initio, starting from the very Lagrangian of QCD, are numerical simulations of the lattice discretized version of this gauge theory. Unfortunately, these lattice QCD simulations can not be performed at finite chemical potentials where they suffer from the yet unsolved sign problem.    
In this situation, different strategies have been followed to develope effective models that could describe the equation of state (EoS) and thus the phase diagram in the entire $T-\mu$ plane, including in particular the critical behavior due to the phase transformation and its possible critical point(s). These models could be calibrated with lattice QCD results in the vicinity of the $T-$axis, for vanishing or small $\mu$ and with constraints from compact star physics at large $\mu$ and $T=0$. 
For the Introduction to the present contribution, we recall two examples for such strategies within the well-known class of two-phase models.
\\
\underline{Maxwell construction:} Separate equations of state for the hadronic phase (HRG) and the pQGP are considered in direct phase equilibrium. As has been pointed out early on (see, e.g., Ref.~\cite{Satz:2018oiz}), this cannot work because of the dominance of the dominant higher number of low-mass degrees of freedom in the pQGP at all temperatures and the absence of confinement - the hadronic world would not exist! Just adding a phenomenological bag pressure works well as a robust thermodynamic confinement mechanism leading to reasonable estimates for the deconfinement transition in the QCD phase diagram. However, by matching these separate HRG and QGP EoS using the Gibbs conditions of phase equilibrium, necessarily a jump in the first derivatives of the thermodynamic potential arises which signals a first order phase transition.
This is in contradiction with the precise results of lattice QCD simulations showing that at least in the domain $\mu/T \lesssim 2$ and $T\gtrsim 130$ MeV the QCD phase transition is a crossover \cite{Bazavov:2018mes}. As there are arguments that also at low temperatures the transition may be a crossover due to symmetries in color superconducting quark matter that entail a continuity of quark(-gluon) and hadron phases \cite{Schafer:1998ef,Wetterich:1999vd}, there may be a second endpoint of the first-order transition at low temperatures \cite{Hatsuda:2006ps} or the transition may even be a crossover all over the QCD phase diagram.
Within the two-phase EoS Maxwell construction scheme one has improved the confining mechanism in the quark matter EoS by a flexible relativistic density functional (RDF) approach that could mimic crossover behavior at small chemical potentials in the sense of a very small latent heat of the first-order transition \cite{Khvorostukin:2006aw}. 
This EoS has been the basis for numerous applications within the three-fluid hydrodynamics simulation approach, see 
\cite{Batyuk:2016qmb,Kozhevnikova:2020bdb} and references therein.
\\
\underline{Interpolation:}
The simple HRG model for the hadronic phase and the bag model for the QGP phase certainly are not suitable for being extended to the very transition region, due to the absence of quark substructure effects in the former and the absence of hadronic correlations in the latter. Therefore, the concept of a phase transition construction directly between these idealized descriptions of the two phases has been replaced by a method that leaves a corridor between these phases which is subject to a separate modeling, an interpolation. This method has been pioneered in \cite{Asakawa:1995zu} for the interpretation of lattice QCD thermodynamics at 
$\mu=0$, but was then further developed also for the case of high $\mu$ and low temperatures in applications to compact star astrophysics, see \cite{Masuda:2015wva,Kojo:2018glx} and references therein.   
A quantitatively reliable interpolation between HRG (even with excluded volume effects mimicking the quark substructure effect of quark Pauli blocking) and a pQGP EoS has been given in  \cite{Albright:2014gva} where an interpolation in the $T-\mu$ plane has been defined by a "switch function" that results in a perfect description of lattice QCD thermodynamics data where they exist and otherwise models a "crossover all over" situation.  At low temperatures this should not be reliable since here the effects of the nuclear liquid-gas phase transition are not captured by a simple HRG (this has been amended, e.g., by 
Vovchenko \cite{Vovchenko:2017cbu}) and it is desireable to model a first-order deconfinement transition with a critical endpoint \footnote{Note that a model with two critical endpoints in the QCD phase diagram can also be obtained when extrapolating a RDF approach to nuclear matter with a $T-$ and $\mu-$dependent excluded hadron volume \cite{Typel:2017vif}.} 
(which has been realised, e.g., in \cite{Plumberg:2018fxo}).
For defining an interpolation with a critical point a method has been suggested in \cite{Nonaka:2004pg} which makes uses the knowledge of the thermodynamic behavior in the vicinity of the critical point of the Ising model.
Such an interpolation approach has been developed further and in the form described in Ref. \cite{Parotto:2018pwx} became a workhorse model for the EoS to be used in simulations of heavy-ion collisions in the framework of the beam energy scan theory
investigations.

The above models, albeit being successful in describing existing lattice QCD data and providing a tool for investigating the consequences of a critical endpoint of the deconfinement transition, are not satisfactory as any microphysical details of the transition itself cannot be addressed by the assumption of a switch function.
In the main section of this contribution we report on first results of elucidating the aspect of hadronic bound state dissociation in 
the transition region that have been obtained within a cluster expansion of the generalized Beth-Uhlenbeck approach to the thermodynamics of nonideal plasmas \cite{Schmidt:1990oyr,Hufner:1994ma,Blaschke:2013zaa,Blaschke:2016fdh}.



\section{Cluster expansion of the Beth-Uhlenbeck approach}
The main idea for unifying the description of the QGP and the HRG phase of low-energy QCD matter is 
the fact that hadrons are 
bound states (clusters) of quarks  and should therefore emerge in a cluster expansion of interacting quark matter
as new, collective degrees of freedom \cite{Bastian:2018wfl}.  
For the total  thermodynamic potential of the model, from which all other equations of state can be derived, 
we make the following ansatz \cite{Blaschke:2020lhm}
\begin{eqnarray}
\Omega_{\rm total}(T;\phi) = \Omega_{\rm QGP}(T;\phi) + \Omega_{\rm MHRG}(T) ~,
\end{eqnarray}
where $\Omega_{\rm QGP}(T;\phi)=\Omega_{\rm PNJL}(T;\phi)+ \Omega_{\rm pert}(T;\phi)$ 
describes the thermodynamic potential of the quark and gluon degrees of freedom with a 
perturbative part $\Omega_{\rm pert}(T;\phi)$ and a nonperturbative mean field part $\Omega_{\rm PNJL}(T;\phi)=\Omega_{\rm Q}(T;\phi)+\mathcal{U}(T;\phi)$ 
that can be decomposed into the quark quasiparticle contribution $\Omega_{\rm Q}(\phi;T)$ and the gluon contribution that is approximated by a (mean field) Polyakov-loop potential $\mathcal{U}(T;\phi)$ \cite{Ratti:2005jh}.
For the MHRG part of the pressure of the model, we have $P_{\rm MHRG}(T)= - \Omega_{\rm MHRG}(T)$
\begin{eqnarray}
P_{\rm MHRG}(T) = \sum_{i = M,B}{P_i(T)}~,
\end{eqnarray}
where the sum extends over all mesonic (M) and baryonic (B) states from the particle data group (PDG), comprising an ideal mixture of hadronic bound and scattering states in the channel $i$ that are described by a Beth-Uhlenbeck formula. 
Then the partial pressure of the hadron species $i$ reads
\begin{equation}
\label{BU}
P_i(T)=\mp d_i\int_0^\infty\frac{dp~p^2}{2\pi^2}\int_0^\infty \frac{dM}{\pi}
	~T \ln \left(1\mp {\rm e}^{-\sqrt{p^2+M^2}/T}\right) \frac{d \delta_i(M;T)}{dM}~,
\end{equation}
where $d_i$ is the degeneracy factor. 
For the phase shift of the bound states of $N_i$ quarks in the hadron $i$ we adopt
the simple model that is in accordance with the Levinson theorem
\begin{equation}
\label{phase}
\delta_i(M;T)=\pi~\Theta(M-M_i)\Theta(M-M_{{\rm thr},i}(T)).
\end{equation}
It describes the occurrence of a bound state at the mass $M_i$ (step-up) and its removal
(step-down) when this mass hits the corresponding continuum threshold $M_i=M_{{\rm thr},i}(T,\mu)$ 
(Mott effect) at the Mott temperature $T_{\rm Mott}(\mu)$.
Inserting (\ref{phase}) into (\ref{BU}) results in 
\begin{equation}
\label{BU-Mott}
P_i(T)=\mp d_i\int_0^\infty\frac{dp~p^2}{2\pi^2}
	~T \ln \left[ \frac{\left(1\mp {\rm e}^{-\sqrt{p^2+M_i^2}/T}\right)}{
	\left(1\mp {\rm e}^{-\sqrt{p^2+M_{{\rm thr},i}(T)^2}/T}\right)}\right]~\Theta(M_{{\rm thr},i}(T)-M_i).
\end{equation}
Now we discuss two examples for the application of the approach.

\subsection{The high-temperature case}
\begin{figure}[htb]
\centerline{%
\includegraphics[width=13.5cm]{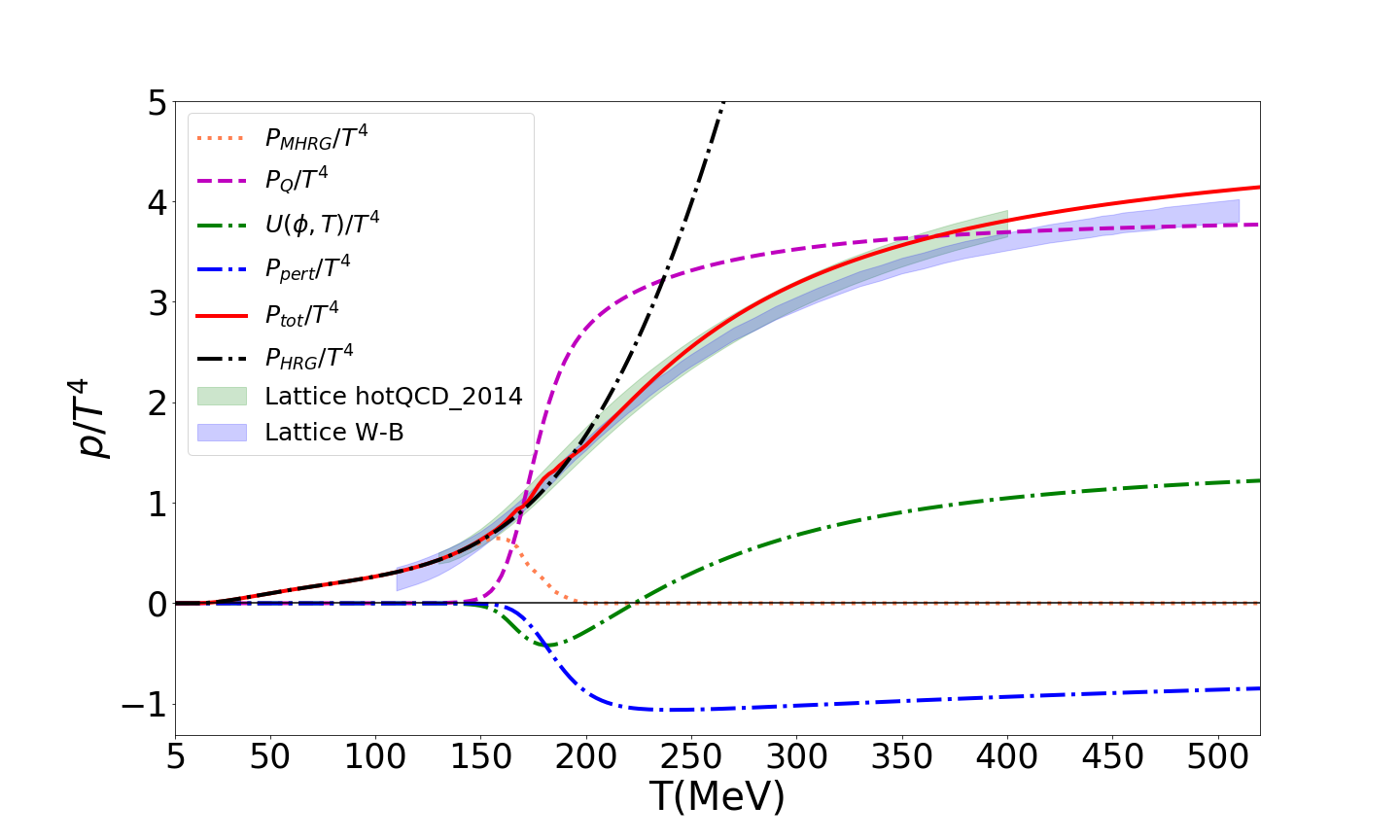}}
\caption{The temperature dependence of the total scaled pressure (red solid line) and it's constituents: MHRG (coral dotted line), quark (dashed magenta line), Polyakov-loop potential $\mathcal{U}(T;\phi)$ (dash-dotted green line), perturbative QCD contribution (dash-dotted blue line) compared to the lattice QCD data: [HotQCD Collaboration] \cite{Bazavov:2014pvz} (green band) and [Wuppertal-Budapest Collaboration] \cite{Borsanyi:2013bia} (blue band). For comparison also the HRG is shown (dash-dotted black line). For details, see Ref.~\cite{Blaschke:2020lhm}.
}
\label{Fig:1}
\end{figure}

In Ref.~\cite{Blaschke:2020lhm} we have applied the cluster decomposition approach to the hadron-to-quark matter transition at finite $T$ and vanishing $\mu$ where we perform a comparison with modern lattice QCD results.
Here we do not solve the gap equation for the dynamical quark mass, but we adopt a linear relation between the threshold masses in the different hadronic channels and the chiral condensate $\Delta_{l,s}(T)$ that we fit to the lattice QCD data.
The chiral symmetry restoration encoded in this behavior triggers the Mott dissociation of all components of the HRG.
For the temperature dependence of the Polyakov loop $\phi$, however, we solve the gap equation that follows from the stationarity of the thermodynamical potential  $ \Omega_{\rm QGP}(T;\phi)$ w.r.t. variation of $\phi$. In this way, the thermodynamic behaviour of all three contributions to the QGP pressure becomes synchronized and results in a smooth behavior.
The temperature dependence of all contributions to the pressure is shown in Fig.~\ref{Fig:1}.

Despite the fact that a simple HRG model alone describes the lattice QCD data well up to $T\sim 200$ MeV, i.e. beyond the pseudocritical temperature $T_c=156.5\pm 1.5$ MeV \cite{Bazavov:2018mes}, the account for the quark substructure of the hadrons in the generalized Beth-Uhlenbeck approach leads to their Mott dissociation in the temperature region 
$150< T[{\rm MeV}]<200$. In this temperature region the quark quasiparticle contribution rises from zero to almost the asymptotic
Stefan-Boltzmann behavior, much steeper than the full result from lattice QCD simulations.
For the very good agreement of our approach with the latter it is important that the contribution from the Polyakov loop potential is negative (like a bag pressure) in the transition region and asymptotically describes the Stefan-Boltzmann contribution of the gluons.
It is well known and confirmed by the lattice QCD data that the behaviour at high temperatures deviates from the Stefan-Boltzmann 
one for quarks and gluons as massless, ideal quantum gases but can be well described by virial corrections in 
$\mathcal{O}(\alpha_s)$ perturbation theory, see \cite{Blaschke:2020lhm} for the details. 

\subsection{The low-temperature case}
\begin{figure}[htb]
\centerline{%
\includegraphics[width=12.5cm]{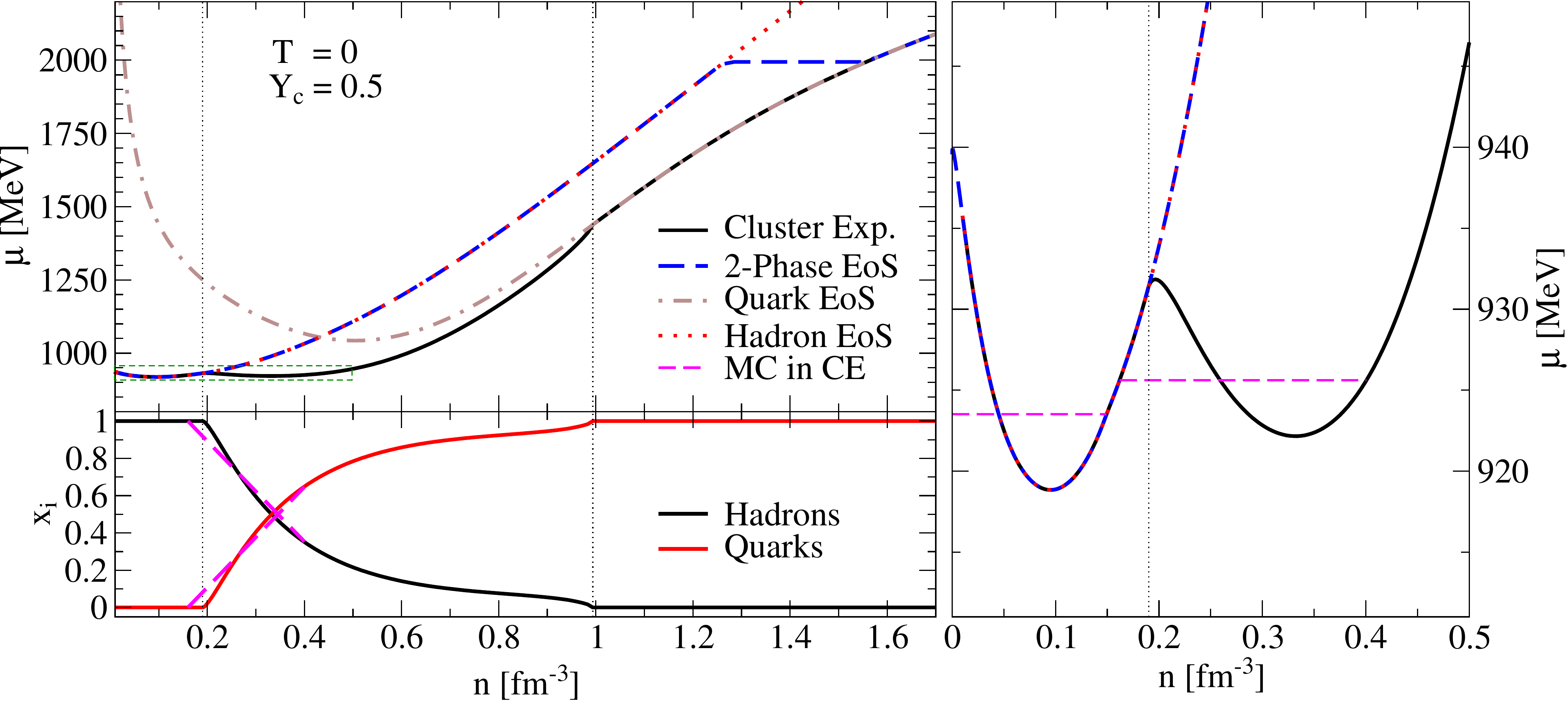}}
\caption{Phase transition from hadronic to quark EoS shown for both the cluster expansion and the two-phase approach.
	The upper left panel shows the chemical potential over density comparing different approaches, while the green dashed box, showing the phase transition of the cluster expansion, is magnified in the right panel. 
	The vertical black thin dotted lines show the region, where both quarks and hadrons exist in the system of cluster expansion.
	The lower left panel shows the concentration of hadrons and quarks from the cluster expansion, where the dashed lines are the modifications due to Maxwell construction. Abridged from \cite{Bastian:2018mmc}.}
\label{Fig:2}
\end{figure}

In the low-temperature case ($T<100$ MeV) that has been discussed in Ref.~\cite{Bastian:2018mmc}, the same ansatz (\ref{phase}) for the hadronic phase shifts has been made, but the number of relevant components in the HRG is reduced to that 
of the nucleons, since no mesons get excited at this low $T$ and we restrict ourselves to chemical potentials below the hyperon threshold. 
With similar arguments the Polyakov-loop potential and perturbative QCD corrections are neglected in this first step
calculation.
An important role, however, plays an effective interaction energy density functional depending on the scalar and vector densities
 \cite{Kaltenborn:2017hus} 
which determine the quasiparticle dispersion relations for the quarks and nucleons and have to be determined selfconsistently by solving corresponding gap equations. 
The result for the density $n=-\partial \Omega/\partial \mu_j$, $j=n,q$; can be inverted and from the Gibbs potential $\mu(n;T)$ one obtains the thermodynamics of the system. The results shown in Fig.~\ref{Fig:2} exhibit two van-der-Waals wiggles which correspond to the liquid-gas transition in nuclear matter and the deconfinement transition, respectively.

\begin{figure}[htb]
\centerline{%
\includegraphics[width=10.5cm]{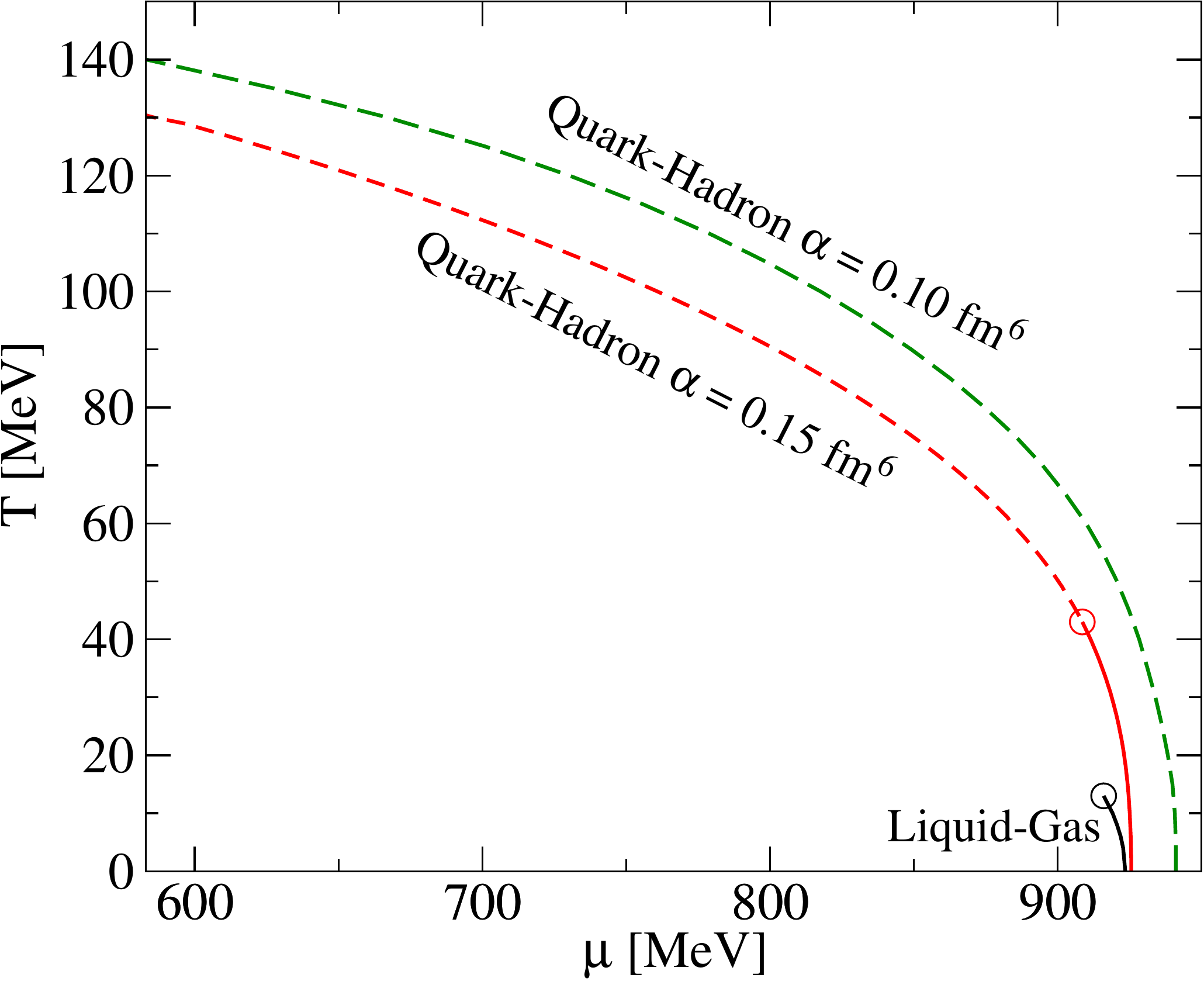}}
\caption{Phase diagram of the cluster expansion with the nuclear liquid-gas phase transition (black) and two parametrizations of quark-hadron transition. The red lines feature a first order phase transition with mixed phase and a critical endpoint at 
$T_\mathrm c ={43}$ MeV and $\mu_\mathrm c = {909}$ MeV. The green dashed line represents a crossover all over scenario without critical endpoint.
The dashed lines show the crossover transition, retrieved from the inflection point in chemical potential over density.
Abridged from \cite{Bastian:2018mmc}.
}
\label{Fig:3}
\end{figure}
In Figure \ref{Fig:3} we summarize these results for the phase diagram in the low-temperature region where two critical endpoints are obtained. A slight change in the model parameter responsible for the screening of the string tension in the density functional leads to a replacement of the second CEP by the case of "crossover all over" for the deconfinement transition. 
This is a remarkable result for such a simple unified model of quark-nucleon matter. 
In a parallel attempt at a unified description of quark-nuclear matter applied in \cite{Marczenko:2020jma} to Astrophysics of compact stars, the aspect of nucleonic parity doubling at the chiral restoration transition has been included and leads to another corner of first-order phase transition at low temperatures in the QCD phase diagram.

\section{Conclusions}
In this contribution we have presented a promising approach to a unified description of HRG and QGP in the QCD phase diagram, whereby the focus was on the hadron dissociation as decisive element for such a nonperturbative marriage. It has been accomplished within a cluster decomposition of the generalized Beth-Uhlenbeck approach and the results of two benchmark calculations have been presented here. They show very good agreement with lattice QCD thermodynamics on the $T-$ axis and the capability of describing a critical endpoint as well as a crossover-all-over situation for the deconfinement transition which would be not possible within  two-phase approaches employing a Maxwell construction of the phase transition.  
In developing the approach further, both regions of the QCD phase diagram shall be described with the same model assumptions and an extension to finite isospin asymmetries as well as inclusion of the parity doubling aspects in the HRG component of the EoS shall be accomplished. These aspects are important preconditions for the application of the approach in simulations of heavy-ion collisions as well as astrophysical phenomena like neutron stars, their mergers and supernovae.

\subsection*{Acknowledgements}
I am grateful to N.-U.F. Bastian, M. Buballa, K. A. Bugaev, K.A. Devyatyarov, A. Dubinin, O. Ivanytskyi, O. Kaczmarek, S. Liebing, 
M. Marczenko, K. Maslov, K. Redlich, G. R\"opke, L. Turko for their collaboration and discussions on this subject. 
This work was supported by the Polish National Science Center under grant No. 2019/33/B/ST9/03059 
and by the Russian Foundation for Basic Research under grant No. 18-02-40137.

\end{document}